\documentclass[iop, twocolumn, tighten]{emulateapj}
\usepackage{amsmath}
\usepackage{color}
\definecolor{redgray}{rgb}{0.2,0.0,0.0}
\definecolor{red}{rgb}{0.8,0.0,0.0}

\newcommand{\bulk}{\mathrm{b}}

\newcommand{\kic}{KIC 12557548\,}
\newcommand{\kicb}{KIC 12557548b\,}

\newcommand{\Qlast}{15}

\newcommand{\dpp}{d_\mathrm{1.5h}}

\newcommand{\Pdep}{P_1 = 22.83 \pm 0.21 \,}
\newcommand{\xuv}{\mathrm{XUV}}

\slugcomment{Published in ApJ Letters, 776:L6 (2013)}
\shortauthors{Kawahara et al.}
\shorttitle{}
\begin{document}
\title{Starspots - Transit Depth Relation of the Evaporating Planet Candidate KIC 12557548b}

\author{Hajime Kawahara\altaffilmark{1}, Teruyuki Hirano\altaffilmark{2}, Kenji Kurosaki\altaffilmark{1}, Yuichi Ito\altaffilmark{1,2}, and Masahiro Ikoma\altaffilmark{1}} 
\altaffiltext{1}{Department of Earth and Planetary Science, The University of Tokyo, 
Tokyo 113-0033, Japan}
\altaffiltext{2}{Department of Earth and Planetary Sciences, Tokyo Institute of Technology, Tokyo 152-8550, Japan}
\email{kawahara@eps.s.u-tokyo.ac.jp}
\begin{abstract}
Violent variation of transit depths and an ingress-egress asymmetry of the transit light curve discovered in \kic have been interpreted as evidences of a catastrophic evaporation of atmosphere with dust ($\dot{M}_p \gtrsim 1 M_\oplus \mathrm{Gyr}^{-1}$) from a close-in small planet. To explore what drives the anomalous atmospheric escape, we perform time-series analysis of the transit depth variation of {\it Kepler} archival data for $\sim$ 3.5 yr. We find a $\sim 30$\% periodic variation of the transit depth with $\Pdep$ days, which is within the error of the rotation period of the host star estimated using the light curve modulation, $P_\mathrm{rot} = 22.91 \pm 0.24$ days. We interpret the results as evidence that the atmospheric escape of \kicb correlates with stellar activity. We consider possible scenarios that account for both the mass loss rate and the correlation with stellar activity. X-ray and ultraviolet (XUV)-driven evaporation is possible if one accepts a relatively high XUV flux and a high efficiency for converting the input energy to the kinetic energy of the atmosphere. Star-planet magnetic interaction is another possible scenario though huge uncertainty remains for the mass loss rate. 
\end{abstract}

\section{Introduction}

Planetary evaporation is one of the most crucial factors that determines the evolution and fate of close-in planets. The atmospheric escapes of hot Jupiters have been extensively investigated from both the extreme ultraviolet observations \citep[e.g.,][]{2003Natur.422..143V,2004ApJ...604L..69V,2010A&A...514A..72L,2012A&A...547A..18E} and theory \citep[e.g.,][]{2003ApJ...598L.121L,2004A&A...419L..13B,2004Icar..170..167Y,2004A&A...418L...1L,2007A&A...472..329E,2007P&SS...55.1426G,2009ApJ...693...23M}. The impacts of evaporation on the evolution of even smaller planets, including super-Earths and super-Neptunes, has been also recognized \citep[e.g.][]{2010A&A...516A..20V,2012MNRAS.425.2931O,2012ApJ...761...59L,2013arXiv1303.3899O,kih}. Recently \citet{2012ApJ...752....1R} (hereafter R12) found violent variation of transit depths and an ingress-egress asymmetry of \kic, which have been interpreted as evidences of a catastrophic evaporation of a small planet \citep[R12,][hereafter PC13]{2013MNRAS.tmp.1542P}. Analyzing the intensity of forward scattering, \citet{2012A&A...545L...5B} and \citet{2012arXiv1208.3693B} found that the dust expected as an occulter of \kic is consistent with sub-micron grains. R12 and PC13 estimated the mass loss rate of \kicb, as $\dot{M}_p \gtrsim 1 M_\oplus \mathrm{Gyr}^{-1}$.  

However, the dominant energy source of the enormous evaporation remains unknown. One proposed possible scenario is that high bolometric energy input at the semi-major axis $a=0.013$ AU itself produces the hydrodynamic escape. PC13 have constructed a radiative-hydrodynamic model of \kicb assuming a Parker-type wind driven by hot planetary surface with the equilibrium temperature $\sim$ 2000 K. They concluded that a planet with mass $M_p \lesssim 0.02 M_\oplus$ can account for the mass loss rate of \kicb. The hydrodynamic escape driven by X-ray and ultraviolet (XUV) radiations as has been discussed for hot Jupiters is another candidate for the energy source. However, an XUV observation of the transit is impossible for \kic with current satellites due to its distance.  

Instead, we focus on the time-series of the transit depth. In this Letter, we reanalyze the {\it Kepler} data of \kic, but for longer period $\sim$ 3.5 yr, than that analyzed by R12. We find the correlation between the transit depth and the modulation of the light curve using starspots. These results indicate that the evaporation of \kicb results from some energy related to the stellar activity, rather than the bolometric one. 

\section{Analysis of Depth Variation}
\begin{figure}[!tb]
  \includegraphics[width=0.99 \linewidth]{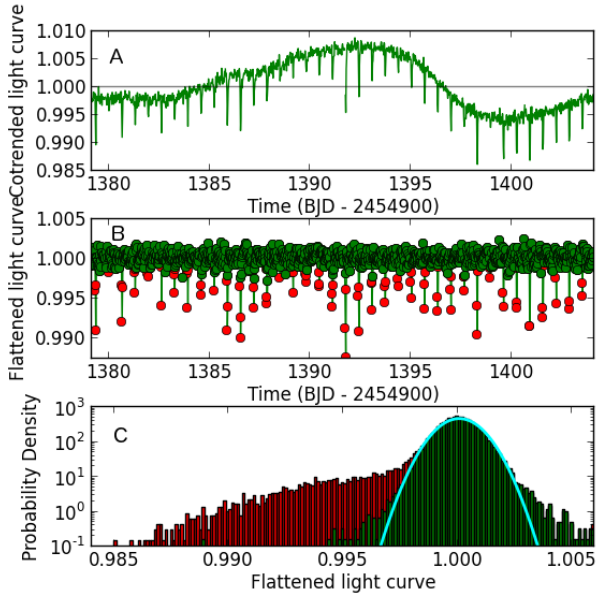}
\caption{Panel A: part of the cotrended light curve of long time cadence of \kic. Panel B: flattened light curve and identified periods (red points) of the transits. Panel C: frequency distribution of the flattened light curve. The red tail corresponds to the transits detected by our method. The rest (green) is well fitted by a normal distribution (cyan). \label{fig:lc}}
\end{figure}

\begin{figure}[!tb]
  \includegraphics[width=0.99 \linewidth]{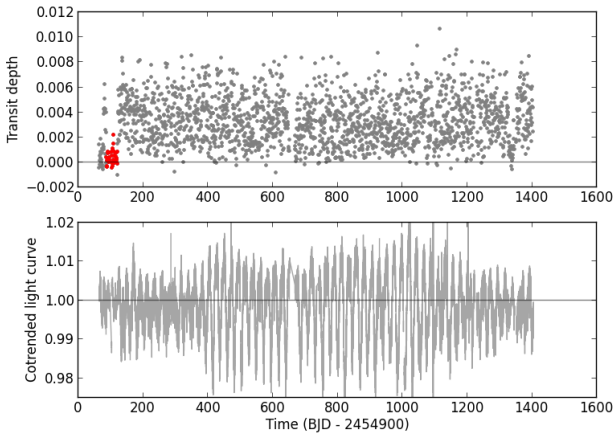}
\caption{Transit depth variation (top) with cotrended light curve (bottom). One can see the period marked by red points in which transits are mostly undetected (see Section 3). \label{fig:tradep}}
\end{figure}

We analyze archived long cadence data of \kic between quarter 1 and \Qlast (Q1-Q15;  $\sim$ 3.5 yr) using the ${\it PyKE}$ package to process the data \citep{2012ascl.soft08004S}. Figure \ref{fig:lc} shows the cotrended long time cadence data of \kic (Panel A) and the flattened light curve (Panel B), extracted by {\it kepcotrend} and {\it kepflatten}. The time series of the transit depth is obtained by averaging three bins (=1.47 hr) around the central time of transit. We adopt the central time as $t_n  = n P_\mathrm{orb} + \mathrm{offset}$, where $P_\mathrm{orb}=0.653558$ days (R12) and the $\mathrm{offset} = 2454900.352921$ days. Our definition of the transit depth is the average of the 3 bins of the flatten light curve, as shown by red points in Figure \ref{fig:lc}(B). As shown in Figure \ref{fig:lc}(C), the distribution of the flatten light curve after removal of the transit periods is well fitted by a normal distribution, at least for the lower side, which verifies that our extraction method could extract almost all of the transit periods. Figure \ref{fig:tradep} shows the time series of the extracted transit depth (top) with the cotrended light curve of the star (bottom). 
 
\subsection{Periodicity of Transit Depth}

\begin{figure}
  \begin{center}
  \includegraphics[width=\linewidth]{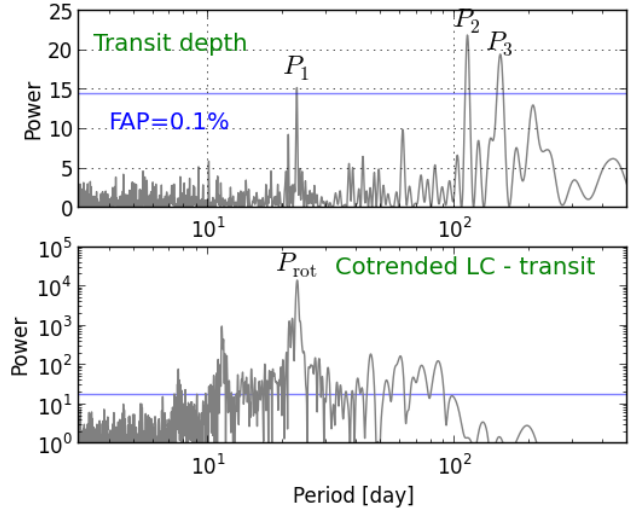}
\caption{Lomb-Scargle periodograms of the depth variation (top) and cotrended light curve without transit duration (bottom). The vertical lines indicate the 0.1 \% level of the FAP.  \label{fig:period}}
  \end{center}
\end{figure}
Figure \ref{fig:period} shows Lomb-Scargle periodograms of the time series of the transit depth. We find three peaks above false alarm probability (FAP) =0.1 \%, $\Pdep$ days,$ P_2 = 112.1 \pm 3.0$ days,  and $P_3 =  152 \pm 7$ days, corresponding to FAP= 0.04 \%, $10^{-5}$ \%, and $6 \times 10^{-4}$ \%. We also compute periodogram of the cotrended light curve excluding the detected transit duration (bottom panel in Figure \ref{fig:period}). We find the most prominent peak at $P_\mathrm{rot} = 22.91 \pm 0.24$ days and its harmonic peaks. 

We also performed high resolution spectroscopy for KIC 12557548 with the High Dispersion Spectrograph on the Subaru telescope. We adopted the standard ``Ra'' setup with the image slicer No.2 ($R\sim 80,000$ and $510-780$ nm). The basic spectroscopic parameters were extracted based on \citet{2002PASJ...54..451T} and \citet{2012ApJ...756...66H}: $T_\mathrm{eff}=4950\pm70$ K, $\log g=3.9\pm 0.2$, $[\mathrm{Fe/H}]=0.09\pm0.09$, and $v\sin i=1\pm1$ km s$^{-1}$. The resultant systematic error in $v\sin i$ is quite uncertain because of the faintness of the target star ($V\sim 16$), but we can safely rule out a large rotational velocity, $v\sin i\gtrsim 4$ km s$^{-1}$, corresponding to a rapid rotation $P_\mathrm{rot} \lesssim 8$ days (we assume $R_\star=0.65 R_\odot$; R12).

On the assumption that this value is the stellar rotation period, the shortest periodicity found in the transit depth variation $\Pdep$ days is consistent with the stellar rotation period. We concentrate on the periodicity of $P_1$ through the rest of the paper. 

\begin{figure*}
\begin{center}
  \includegraphics[width=0.7\linewidth]{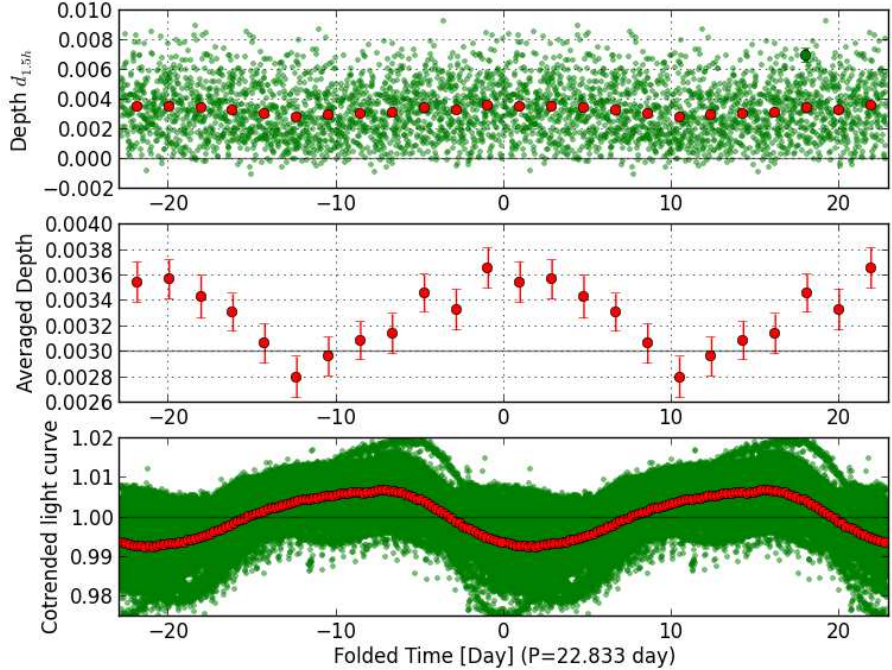}
\caption{Top panel: the transit depth variation folded by $\Pdep$ days. A typical statistical error of the depth is shown by a green bar. Red bars are the transit depth averaged over bins. Middle panel: the mean value of the binned data of the top panel (an expanded version of the top panel around the red error bars). Bottom panel : the folded cotrended light curve of \kic. \label{fig:lcfold}}
\end{center}
\end{figure*}

We fold the transit depth variation with $P_1$ (the top and middle panels in Figure \ref{fig:lcfold}). The mean value of binned data has $\sim$ 30 \% variation. We also fold the cotrended light curve with $P_1$ (the bottom panel) and find that the folded cotrended light curve is negatively correlated with the folded depth variation. A large starspot can survive for many years \citep{2005LRSP....2....8B}. The 2 \% variation of the folded light curve can be interpreted as long term variation due to a large starspot associated with a local active area. Hence our interpretation of the anti-correlation is that the planet tends to make deeper occultation when facing the large starspot.
 
In general, stellar visibility depends on competition among starspots and faculae. In the case of the Sun, spot modeling \citep{2007A&A...464..741L} predicts that a single active area decreases the visibility only for the angle between the line-of-sight and the local active area on the stellar surface $\phi \lesssim 45^\circ$ due to the large faculae-to-spots ratio $Q=9$. However, for several stars with large flux modulation, smaller values of $Q=1-4.5$ were estimated from the spot modeling and faculae for these stars do not contribute significantly to the visibility  \citep{2009A&A...493..193L,2009A&A...506..255L,2010A&A...520A..53L}. Throughout the rest of the paper, we assume that the local active area decreases stellar visibility.

The stellar visibility variation results in the apparent variation of the measured depth via normalization. \citet{2011A&A...526A..12D} considered this effect for different epochs of stellar activity (a $\sim$ 1 year interval). We use a similar formalism although we consider this effect during one stellar rotation. The apparent variation of the depth $\delta_d \equiv (d_{v}-d_{o})/d_{o}$ is related to that of the visibility during one revolution, $\delta_F \equiv (F_{v}-F_{o})/F_{o}$ as 
\begin{eqnarray}
\delta_d = \alpha \delta_F,
\end{eqnarray}
where $d_v$ and $d_o$ are the measured depths when the active area is located on the side visible to us and the opposite side, $F_v$ and $F_o$ are the visibilities for the two side. Denoting the average surface brightness of photosphere by $f_v$ and $f_o$ for the two sides and that along the trajectory of transit by $\tilde{f}_{v}$, $\tilde{f}_{o}$, we can express 
\begin{eqnarray}
\delta_d = \frac{(\tilde{f_v}/f_v)}{(\tilde{f_o}/f_o)} -1 = \frac{\delta_{\tilde{f}}/\delta_f -1}{1 + \delta_F} \delta_F,
\end{eqnarray}
where $\delta_f \equiv (f_{v} - f_{o})/f_{o} = \delta_F$ and $\delta_{\tilde{f}} \equiv (\tilde{f}_{v} - \tilde{f}_{o})/\tilde{f}_{o}$. Assuming that the spot crossing is negligible ($|\delta_{\tilde{f}}/\delta_{f}|<1$) and $1+\delta_F \sim 1$, we obtain $-2 \lesssim \alpha \le 0$. If the surface brightness outside the spot areas is homogeneous, we obtain $\alpha=-1$. In our case, we expect the 0-4 \% variation of $\delta_d$ to be anti-correlated to the stellar visibility. Since the observed variation is $\delta_d \sim$ 30 \%, we conclude that the periodicity of the depth variation synchronized with the stellar rotation is not a false positive. 


 \citet{2013arXiv1307.6959V} found seasonal variation of transit depth for HAT-P-7 b due to instrumental systematics. They did not report any other periodicity $<$ 1 yr. Since the periodicities $P_2$ and $P_3$ are larger than period of a quarter, instrumental systematics should be considered in further studies. 

\section{Interpretation}

To consider possible scenarios of evaporation, two important constraints from observation, the mass-loss rate $\dot{M}_p$ and the planet radius $R_p$, should be taken into account. The mass-loss rate was constrained by both R12 and PC13 in similar but slightly different ways, $\dot{M}_p \sim 1 M_\oplus \mathrm{Gyr}^{-1}$ and $\dot{M}_p > 0.1-1 M_\oplus \mathrm{Gyr}^{-1}$. Here we summarize the essential part of their derivation. The $\dot{M}_p$ can be estimated using the total cross section of transit $S=\pi R_\star^2 d = 7 \times 10^{19} \mathrm{cm^2}$ (we adopt $d=0.01$), dust density $\rho_\mathrm{dust}$, typical length of optical path $\Delta L$ and typical time scale $\Delta t$, 
\begin{eqnarray}
\dot{M}_p &\gtrsim& (1+\eta) \rho_\mathrm{dust} S \Delta L/\Delta t \nonumber \\
&=& 1 \times (1+\eta) \left( \frac{\rho_\mathrm{dust}}{1 \mathrm{g/cm^3}} \right)  \left( \frac{\Delta L}{0.1 \mu \mathrm{m}} \right) \left( \frac{\Delta t}{1 \mathrm{hr}} \right)^{-1} \, M_\oplus/\mathrm{Gyr} \nonumber \\ 
\end{eqnarray}
 where $\eta$ is the gas-dust ratio. R12 adopted $\Delta t = R_\star v = 1.3$ hr, assuming the typical velocity of outflow $v=10$ km/s, $\rho_\mathrm{dust}=3 \mathrm{g/cm^2}$, and $\Delta L=0.2 \mu \mathrm{m}$ as a typical grain size, which yields $\dot{M}_p \sim 1 (1+\eta) M_\oplus/\mathrm{Gyr}$. PC13 uses an orbital period of $\Delta t = P_\mathrm{orb} = 15.7$ hr, $\rho_\mathrm{dust}=1 \mathrm{g/cm^2}$, $\Delta L=0.5 \mu \mathrm{m}$, which results in $\dot{M}_p \sim 0.4 (1+\eta) M_\oplus/\mathrm{Gyr}$. 

The grain size estimated from the forward scattering supported $s \sim 0.1 \mu \mathrm{m}$ \citep{2012A&A...545L...5B,2012arXiv1208.3693B}. We computed the auto correlation of the $\dpp$ and found a flat correlation for a time interval, $\tau <10 $ days. It supports $\Delta t < P_\mathrm{orb}$. Assuming that the dust density is an order of $\mathrm{g/cm^2}$, we use $\dot{M}_p > 1 M_\oplus/\mathrm{Gyr}$ as a fiducial constraint of the mass loss rate. 

Another constraint is the upper limit of $R_p$. \cite{2012A&A...545L...5B} estimated the 1 $\sigma$ upper limit of the maximum planetary radius as $R_p < 1.2 R_\oplus$ translating photometric accuracy of the phase folded light curve of no-transit-like feature (around the period in red in Figure \ref{fig:tradep}) into the radius. 

Interpreting the starspots-depth relation in a straight forward fashion, the atmospheric escape rate should correlate with the area of the active area on the stellar surface. Because the optical flux does not change much ($\sim$ 2\%) and is negatively correlated with the transit depth, the bolometric radiation is unlikely to be a source of the relation. 

\subsection{XUV evaporation \label{ss:xuv}}

Strong XUV radiation is often related to stellar spots since the XUV flares originate from magnetic loops. The X and/or UV periodicity with rotation were found in several stars \citep[e.g.][]{2010ApJ...722..343D,2013A&A...552A...7S}. While the sunspots exhibit much smaller variation than \kic, an order-of-magnitude of variability in solar X-ray has been observed over a month \citep[e.g.][]{1998ASPC..154..223S,2004A&ARv..12...71G}. We found a sudden 6 \% increase in the flux during 30 minutes similar to the flare in the short cadence data. Such a strong flare will create strong XUV variation in a timescale of hours, which might account for the transit depth variation in every transit.  

However, R12 questioned whether the input energy is sufficient to account for the mass loss rate. We revisit this problem using the theoretical mass-radius relation of sub/super-Earths with several different compositions. Because of the lack of XUV observation, we manage to estimate the XUV flux of \kic from the X-ray flux - rotation relation, which provides $L_\mathrm{X} = 10^{27-28.5}$ erg/s for $P_\mathrm{rot} = 23$ days \citep{2003A&A...397..147P}. Considering the fact that \kic has a very large active area that creates 2 \% continuous flux variation over 3.5 yr, we adopt $L_\xuv/L_{\mathrm{solar}} = 10^{-5}$, corresponding to $L_\xuv \sim 10^{28.5}$ erg/s.

In the energy-limit regime \citep{1981Icar...48..150W,2003ApJ...598L.121L,2007P&SS...55.1426G,2007A&A...461.1185L}, the mass loss rate can be estimated as 
\begin{eqnarray}
\dot{M}_p(R_p) &=& \epsilon \frac{\pi r_1^2 F_\xuv}{G M_p/r_0} \frac{1}{K_\mathrm{tide}},
\label{eq:energylimited}
\end{eqnarray}
where $r_0$ is the radius where gas is bound to the planet, $r_1$ is the radius where the bulk of incoming XUV flux is absorbed, $\epsilon$ is the efficiency for converting the input energy to the kinetic energy of the atmosphere. Because the base level of the XUV deposition $r_1$ is under discussion \citep[e.g.][]{2004A&A...419L..13B,2007P&SS...55.1426G,2009ApJ...693...23M}, we adopt the radius where optical depth $\sim 1$ conservatively, i.e. $r_0=r_1=R_p$. 

A correction factor for the Roche lobe effect, $K_\mathrm{tide}$,  \citep{2007A&A...472..329E} is given by,
\begin{eqnarray}
K_\mathrm{tide}  = 1 - \frac{3}{2} \left( \frac{R_p}{R_H} \right)  + \frac{1}{2}  \left( \frac{R_p}{R_H} \right)^3,
\end{eqnarray}
where $R_H$ is the Hill radius, 
\begin{eqnarray}
R_H \sim 3.4 \left(\frac{M_p}{M_\oplus} \right)^{1/3} \left(\frac{M_\star}{0.7 M_\odot} \right)^{-1/3} R_\oplus.
\end{eqnarray}

On the basis of \citet{kih}, we compute the mass-radius relation of planets composed of 100 \% water, 50\% water+50\% rocky core, 10\% water+90 \%rocky core (volatile-rich planets), 100 \% rock, and 100 \% iron assuming the equilibrium temperature of \kicb ($\sim$ 1800 K). We neglect the contribution of a thin atmosphere, which should be evaporating, for the 100 \% rock and iron planets to compute the relation. The interior structure of the planet is computed with equations of hydrostatic equilibrium, mass conservation, and energy conservation \citep[e.g.][]{1990sse..book.....K} and the equation of state. A detailed description is found in \citet{kih}.

\begin{figure*}[!tb]
  \includegraphics[width=0.49 \linewidth]{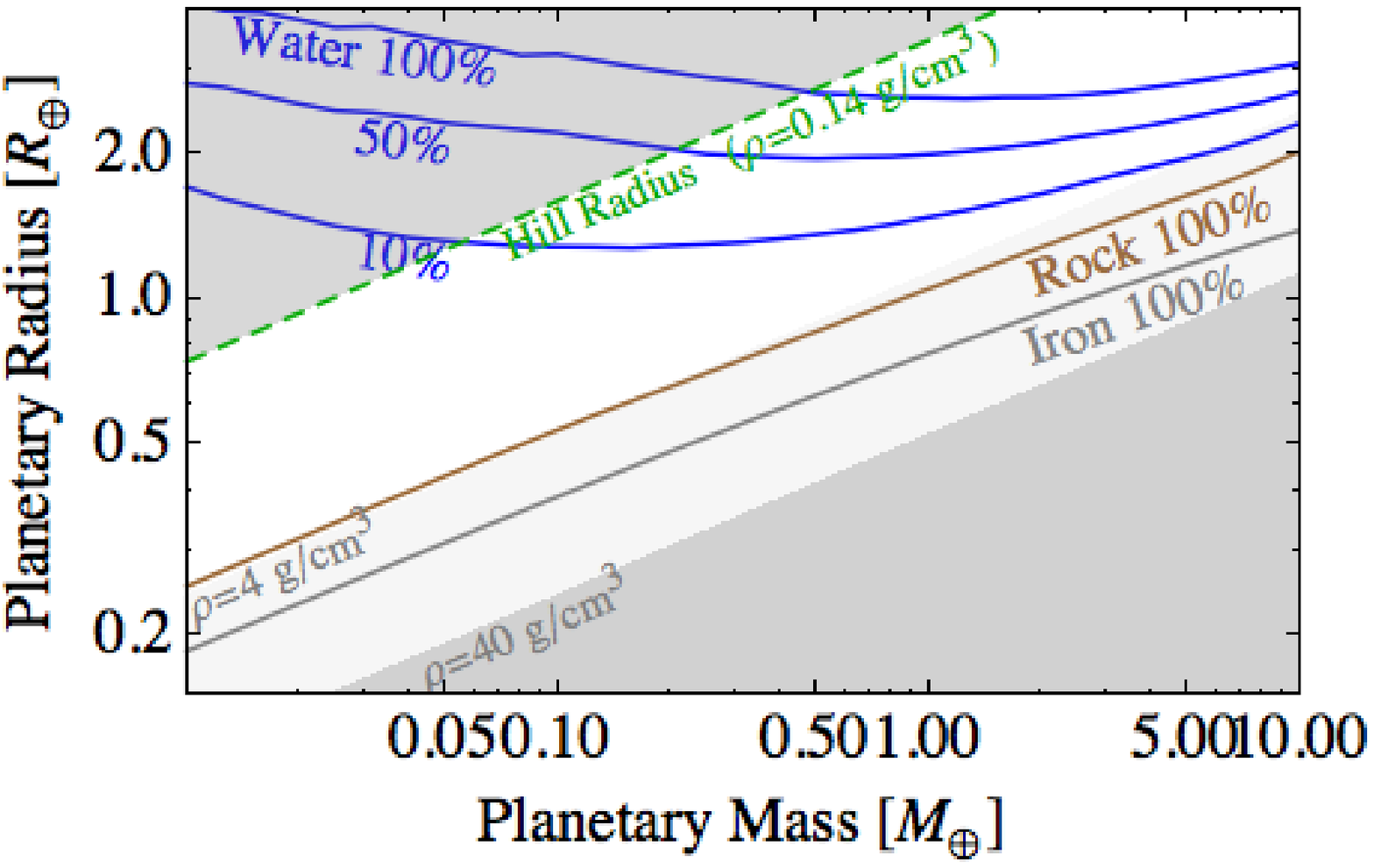}
  \includegraphics[width=0.49 \linewidth]{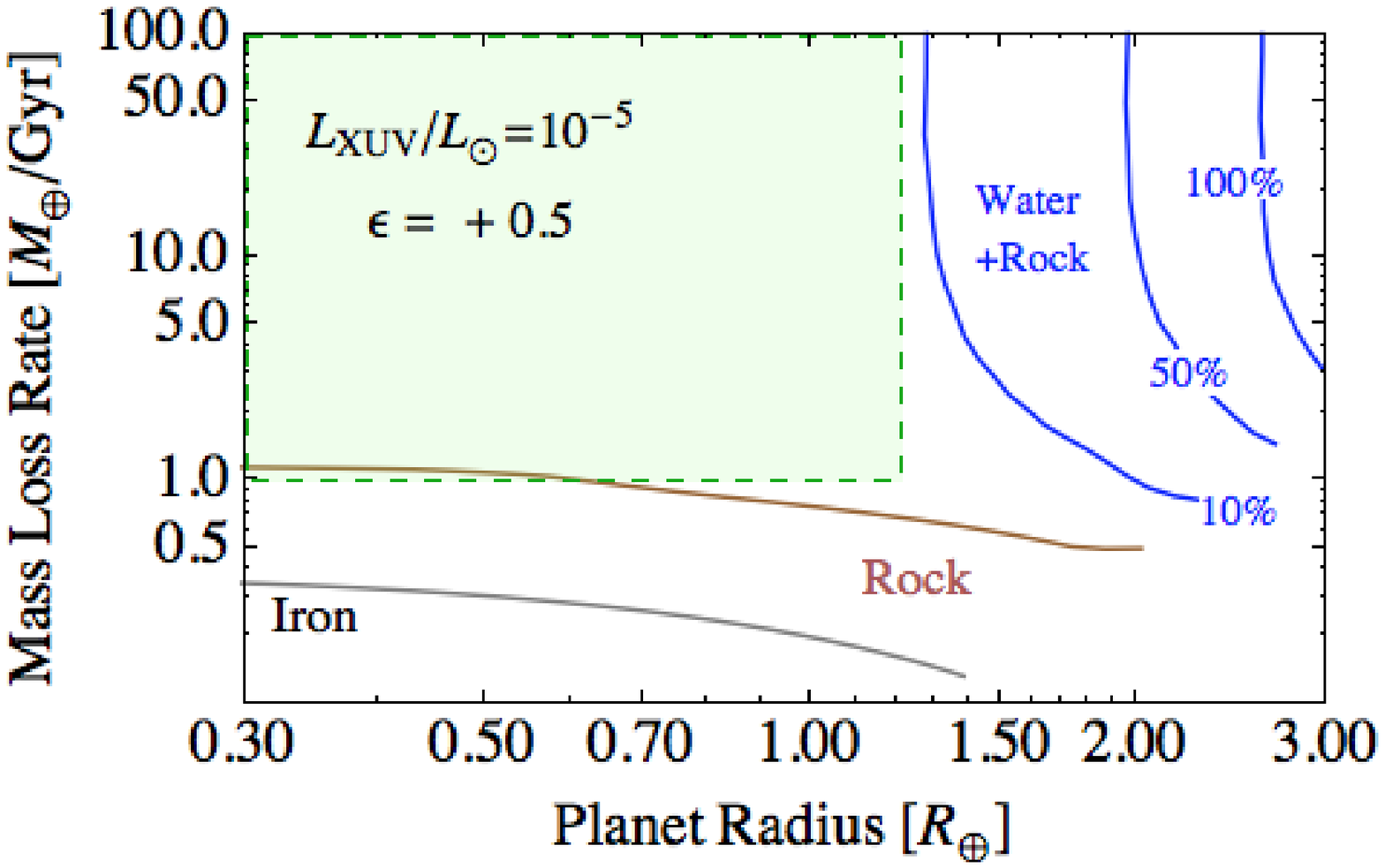}
\caption{The mass-radius relation (left) and $\dot{M}_p$ in the energy-limited regime assuming the XUV driven escape (right). In the left panel, a green dashed line indicates the hill radius assuming $M_\star=0.7 M_\odot$ and $a=0.013$ AU. In the right panel, fiducial  constraints derived by the minimum transit depth and the mass loss estimate are shaded by green. We assume relatively high value of the XUV flux $L_{\xuv} = 10^{-5} L_\odot$ and high efficiency $\epsilon = 0.5$. \label{fig:mlr}}
\end{figure*}

Figure \ref{fig:mlr} (left) shows the mass-radius relation of these planets. We find that the radius of volatile-rich (the water fraction being $\gtrsim$ 10 \%)  planets cannot be smaller than the upper limit of planetary radius $\sim 1.2 R_\oplus$ because the gravity cannot bind the atmosphere. Therefore \kicb is unlikely to be a volatile-rich planet. Figure \ref{fig:mlr} (right) shows the $\dot{M}_p$ expected in the energy-limit case, assuming the relatively high flux $L_\xuv = 10^{-5} L_\odot$ and an assumption for the efficiency of $\epsilon=0.5$. The density of the planet composed of iron is too dense to obtain the mass loss rate needed. The $\dot{M}_p$ of the rocky planet is $\sim 1 M_\oplus$/Gyr. The Roche lobe effect increases $\dot{M}_p$ to two times larger than that without the tide. R12 discussed that the XUV luminosity is at least a factor of 10 too low compared to the luminosity needed assuming $\epsilon=0.1$. The Roche lobe effect and the assumption of $\epsilon$ are likely to cause the difference by one order-of-magnitude in $\dot{M}_p$ between R12 and us. Hence, a rocky planet is possible as a candidate for \kicb.  

One might consider that volatile components on the rocky planet increase the planetary effective radius and $\dot{M}_p$. We derive the surface pressure needed to increase the transit radius $R_t$ to $\chi$ times of the radius of the silicate core $R_b$ ($R_t = R_\bulk + h = (1+\chi) R_\bulk$) though the observed transit radius is not identical to the effective radius for equation (\ref{eq:energylimited}). We adopt $\chi=0.5$, i.e. $\sim 2$ times larger cross section. The atmospheric height from the ground to the point where the optical depth through the planetary limb is unity is estimated by assuming an isothermal atmosphere with temperature $T$ and a power-law dependence of the opacity on pressure \citep[$\kappa \propto P^\gamma$;][]{2010ApJ...712..974R}, 
\begin{eqnarray}
\label{eq:h}
h = \frac{H}{\gamma + 1} \log{\left( \frac{\tau_\bulk}{\tau(r)} \right)}
\end{eqnarray}
where $H=k T/\mu m_H g$ is the pressure scale height, $P_\bulk$, $\mu$ and $m_H$ are the surface pressure at the lower boundary, mean-molecular weight and proton mass. The vertical optical depth $\tau(r)$ is linked to the optical depth of the chord \citep[e.g.][]{2007ApJ...661..502B,2008ApJS..179..484H,2010ApJ...712..974R} as 
\begin{eqnarray}
\label{eq:con}
\tau_t(r) \approx \sqrt{\frac{2 \pi (\gamma+1) r}{H}} \tau(r).
\end{eqnarray}
It yields
\begin{eqnarray}
\label{eq:hhh}
h = \frac{H}{\gamma + 1} \log{\left( \sqrt{\frac{2 \pi (h + R_\bulk)}{H (\gamma+1)}} \frac{\kappa_b P_b}{g}  \right)},
\end{eqnarray}
where $\kappa_b$ is the visible opacity at the surface. Then we obtain
\begin{eqnarray}
P_b &=& \sqrt{\frac{\gamma+1}{\chi+1}} \frac{g}{\kappa_b} \frac{e^{\chi(\gamma+1) \lambda_b}}{\sqrt{2 \pi \lambda_b}} \nonumber \\
 &\sim& 10^{0.4 \chi(\gamma+1) \lambda_b - \Delta} \left( \frac{\kappa_b}{1 \, \mathrm{cm^2/g}} \right)^{-1} \mathrm{[bar]} 
\end{eqnarray}
 where $\lambda_b \equiv G M \mu m_H/k T R_b$ is the escape parameter and $\Delta \sim 4.5$ for $M=(0.01-1) M_\oplus$ and $\rho=4 \,\mathrm{g/cm^3}$. For the Na atmosphere, which is a possible dominant component of the atmosphere of the rocky planets \citep[e.g.][]{2011ApJ...742L..19M}, $\overline{\kappa_b} \sim 10^2 (P/\mathrm{1 bar})^1 $ for $10^{-2}-10^{3}$ bar, where we computed $\kappa_b(\lambda)$ based on Atomic Line Data \citep{1995KurCD..23.....K} with the method by \citet{2001ApJ...547.1040P} and took the logarithm average of $\kappa_b(\lambda)$ (see Eq. [\ref{eq:hhh}]) for the {\it Kepler} band.  Adopting $\mu=30$, we obtain $P_b \sim 10$ bar and $1$ bar for $M=0.1 M_\oplus$ ($\lambda_b \sim 24$) and $M=0.01 M_\oplus$ ($\lambda_b \sim 5$), which are much higher than the equilibrium vapor pressure of the olivine, $10^{-5}-10^{-2}$ bar for 2000-2500 K \citep{2013MNRAS.tmp.1542P} \citep[see also][]{2011ApJ...742L..19M}. Thus the Na atmosphere vaporized at the equilibrium temperature is unlikely to increase the cross section of $\xuv$. Though we assumed the equilibrium temperature as the surface temperature for simplicity, further study of the atmospheric structure over magma ocean is needed to derive the realistic contribution of the volatile components.
 
\subsection{Energetic Electron by Magnetic Reconnection}
Star-planet magnetic interaction is another possible scenario for a catastrophic evaporation that correlates with stellar activity. Recently \citet{2013arXiv1307.2341L} proposed a new scenario in which energetic electrons released by magnetic reconnection at the boundary of the planetary magnetosphere inject kinetic energy into the planetary atmosphere. The power released by the reconnection per area, corresponding flux, is estimated \citep{2012A&A...544A..23L, 2013arXiv1307.2341L} as 
\begin{eqnarray}
F_\mathrm{mag} \equiv \frac{P_\mathrm{mag}}{\pi R_p^2} &=& \frac{B_0^2}{2 \mu_B} \sqrt{\frac{G M_\star}{a}} \left(\frac{B_p}{B_0}\right)^{2/3}  \left(\frac{R_\star}{a}\right)^{4s/3} 
\end{eqnarray}
where $\mu_B$ is the magnetic permeability of the plasma , $B_0$ and $B_p$ are the magnetic field at the pole of a star and a planet. If adopting $B_0=10$ G, $B_p=1$ G, and $s=2$, we obtain $4 \pi a^2 F_\mathrm{mag} = 5 \times 10^{-5} \, (\mu_B/\mu_0)^{-1} L_\odot $, ($\mu_0$ is the vacuum permeability). Since this value is larger than our fiducial value of $L_\xuv = 10^{-5} \, L_\odot$ in Section \ref{ss:xuv}, it is possible to account for the mass loss rate.


\section{Summary}
We performed a time-series analysis of the transit depth of \kicb. We found that the transit depth variation has three values of specific periodicity above FAP=0.1 \% and that the shortest one $\Pdep$ days is consistent with the stellar rotation period estimated by visibility modulation. We interpreted these results as evidence that the evaporation is related to stellar activity.

Using the theoretical mass-radius relation with tide, we showed that the massive escape of a rocky planet driven by XUV is possible if assuming high XUV flux and high efficiency. The energetic electrons released by the magnetic reconnection is another possible scenario. Though we could not identify any plausible scenario to account for the $\sim 1 \dot{M}_\oplus$/Gyr evaporation correlated with stellar activity, future XUV observations will enable us to understand what type of energy evaporates close-in small planets, if transiting planet surveys such as {\it Transiting Exoplanet Survey Satellite} find nearby similar systems.   

We thank Yoichi Takeda for helping with the spectroscopic analysis. This work is supported by the Astrobiology Project of the CNSI, NINS (AB251009) and by Grant-in-Aid for Scientific research from JSPS and from the MEXT (Nos. 25800106, 25400224, and 25-3183).


\end{document}